\documentclass[pre,twocolumn,superscriptaddress]{revtex4-1}
\usepackage{graphicx}
\usepackage{amsmath,amssymb}

\newcommand{\myav}[1]{\langle #1\rangle}

\newcommand{\smallfrac}[2]{{\textstyle{\frac{#1}{#2}}}}
\newcommand{\half}{\smallfrac{1}{2}}

\newcommand{\myvec}[1]{{\mathbf{#1}}}
\newcommand{\rvec}{\myvec{r}}
\newcommand{\kvec}{\myvec{k}}

\newcommand{\kB}{k_\mathrm{B}}
\newcommand{\kT}{\kB T}
\newcommand{\lB}{l_{\mathrm{B}}} 
\newcommand{\kD}{\kappa_{\mathrm{D}}}
\newcommand{\rsq}{\myav{r^2}}
\newcommand{\rhoz}{\rho_z}

\newcommand{\nm}{\mathrm{nm}}
\newcommand{\M}{\mathrm{M}}

\newcommand{\latin}[1]{{\itshape #1}}
\newcommand{\eg}{\latin{e.\,g.}}
\newcommand{\ie}{\latin{i.\,e.}}
\newcommand{\etal}{\latin{et al.}}
\newcommand{\etc}{\latin{et\,c.}}

\newcommand{\Eqref}[1]{Eq.~\eqref{#1}}

\DeclareMathOperator{\erf}{erf}

\begin{document}

\title{Screening properties of four mesoscale smoothed charge models, \\ 
with application to dissipative particle dynamics}

\author{Patrick B. Warren}

\email{patrick.warren@unilever.com}

\affiliation{Unilever R\&D Port Sunlight, Quarry Road East, Bebington,
  Wirral, CH63 3JW, UK.}

\author{Andrey Vlasov}

\affiliation{Department of Chemistry, St. Petersburg State University,
  26 Universitetsky prosp., 198504 St. Petersburg, Russia.}

\date{accepted version -- February 2014}

\begin{abstract}
We extend our previous study [J.~Chem.~Phys.~{\bf138}, 204907 (2013)]
to quantify the screening properties of four mesoscale smoothed charge
models used in dissipative particle dynamics.  Using a combination of
the hypernetted chain integral equation closure and the random phase
approximation, we identify regions where the models exhibit a
real-valued screening length, and the extent to which this agrees with
the Debye length in the physical system.  We find that the second
moment of the smoothed charge distribution is a good predictor of this
behaviour.  We are thus able to recommend a consistent set of
parameters for the models.
\end{abstract}

\maketitle

\section{Introduction}
Dissipative particle dynamics (DPD) has attracted much interest as a
simulation method for soft condensed matter, including charged systems
such as ionic surfactants and water-soluble polyelectrolytes which are
of widespread practical importance \cite{FS02, NMW03, Gro03}.  In such
systems, modelling the electrostatic interactions can be done
implicitly, for example with the Poisson-Boltzmann equation, or
explictly by incorporating charged particles.  In the latter case, in
DPD it is essential to smooth the point charges into charge clouds
since this replaces the $1/r$ divergence of the Coulomb law as
$r\to0$ (where $r$ is the center-center separation) by a smooth
cutoff, thus ensuring thermodynamic stability according to a theorem
by Fisher and Ruelle \cite{MEF66}.

The precise form of the charge smoothing is often tuned to the choice
of numerical algorithm and a consensus on the best approach has yet to
emerge.  At least four different smoothing methods have been suggested
in the literature.  Groot introduced a particle-particle particle-mesh
(P$^3$M) method with linear charge smoothing \cite{Gro03}.  Later
Gonz\'alez-Melchor \etal\ examined an Ewald-based method with
exponential charge smoothing \cite{GMV+06}.  Most recently we have
studied a related Ewald method with Gaussian charge
smoothing \cite{WVA+13}. This last choice connects with recent work on
the so-called ultrasoft restricted primitive model
(URPM) \cite{CHK11a, CHK11b, NHK12}.  Finally, in the context of the
URPM, a Bessel smoothed charge model has been introduced to complement
the Gaussian case \cite{WM13}.

In the present work we extend the study in Ref.~\onlinecite{WVA+13} to
identify regions where the models exhibit a real-valued screening
length, and the extent to which this agrees between models and with
the Debye length in the physical system.  The problem will be
addressed using liquid state integral equation theory \cite{HM06}.  By
focussing on the screening properties of the supporting electrolyte,
we eliminate the need to consider explicit mesoscopic objects such as
polymers and surfactants.

\section{Models}
We first set out the generic DPD electrolyte model which underpins the
rest of the discussion.  More details and justification for particular
parameter choices can be found in previous work \cite{WVA+13}.  We
represent the electrolyte as a multicomponent soft sphere fluid of DPD
particles, containing positive and negative ions of valencies $z_\pm$ at
densities $\rho_\pm$, and a neutral ($z_0=0$) solvent species at a
density $\rho_0$.  The total ion density is $\rho_z=\rho_++\rho_-$, and
the total overall density is $\rho=\rho_0+\rho_z$; these are sufficient to
specify the state point since charge neutrality implies
$z_+\rho_++z_-\rho_-=0$.  The aforementioned charge clouds are
centered on the DPD particles which represent the ions.

\begin{table*}
\begin{ruledtabular}
\begin{tabular}{lccccc}
charge type & $\varrho(r)$ & $g(k)=\tilde\varrho^2$ & $f(r)$
 & $\beta u_0$ & $\rsq$ \\[3pt]
\hline\\[-9pt]
linear\footnote{It is implied that $\varrho(r)=0$ for $r>R$} & 
$\displaystyle \frac{3}{\pi R^3}\Bigl(1-\frac{r}{R}\Bigr)$%
 &
$\displaystyle\frac{144[2-2\cos(kR)-kR\sin(kR)]^2}{(kR)^{8}}$ &
---\footnote{An approximate closed form
expression for $f(r)$ can be found in Appendix A in
Ref.~\onlinecite{Gro03}.} & 
$\displaystyle \frac{52\,\lB}{35\,R}$ & 
$\displaystyle \frac{2}{5}R^2$ \\[15pt]
exponential & 
$\displaystyle \frac{e^{-2r/\lambda}}{\pi\lambda^3}$ &
$\displaystyle \frac{1}{(1+{k^2\lambda^2}/{4})^{4}}$ &
$\displaystyle1-e^{-2r/\lambda}\Bigl(
1+\frac{11r}{8\lambda}+\frac{3r^2}{4\lambda^2}+\frac{r^3}{6\lambda^3}\Bigr)$ &
$\displaystyle \frac{5\lB}{8\lambda}$ & 
$\displaystyle 3\lambda^2$ \\[12pt]
Gaussian &
$\displaystyle \frac{e^{-r^2/2\sigma^2}}{(2\pi\sigma^2)^{3/2}}$ &
$\displaystyle {e^{-k^2\sigma^2}}$ &
$\displaystyle \erf\Bigl(\frac{r}{2\sigma}\Bigr)$ &
$\displaystyle \frac{\lB}{\sigma\sqrt{\pi}}$ &
$\displaystyle 3\sigma^2$ \\[15pt]
Bessel & 
$\displaystyle \frac{1}{2\pi^2\sigma^2
  r}K_1\Bigl(\frac{r}{\sigma}\Bigr)$ &
$\displaystyle \frac{1}{1+k^2\sigma^2}$ &
$\displaystyle 1-e^{-r/\sigma}$ & 
$\displaystyle \frac{\lB}{\sigma}$ & 
$\displaystyle 3\sigma^2$ \\[9pt]
\end{tabular}
\end{ruledtabular}
\caption[?]{Properties of the four smoothed charge models: charge
  distribution $\varrho(r)$, auxiliary function $g(k)$, smoothing
  function $f(r)$ modifying the Coulomb law in \Eqref{eq:u}, overlap
  energy $\beta u_0$, and second moment of charge distribution
  $\rsq$.\label{T1}}
\end{table*}

The fluid particles interact by pair-wise short range soft repulsions
and long range electrostatics, with an interaction potential
\begin{equation}
\beta U_{\mu\nu}(r)=\phi(r)+z_\mu z_\nu \frac{\lB}{r}\,f(r)\,.
\label{eq:u}
\end{equation}
Here $\mu$, $\nu$, labels the species type, and $\beta=1/\kT$ is the
inverse temperature measured in units of Boltzmann's constant.  The
first term in \Eqref{eq:u} is a short range soft repulsion, which
for simplicity we take to be the same for all species.  For
present purposes we do not need to specify $\phi(r)$, other than to
note that $\phi(r)=0$ for $r>r_c$ where $r_c$ represents the size
(radius) of the DPD particles.

\begin{figure}
\begin{center}
\includegraphics[clip=true,width=3.0in]{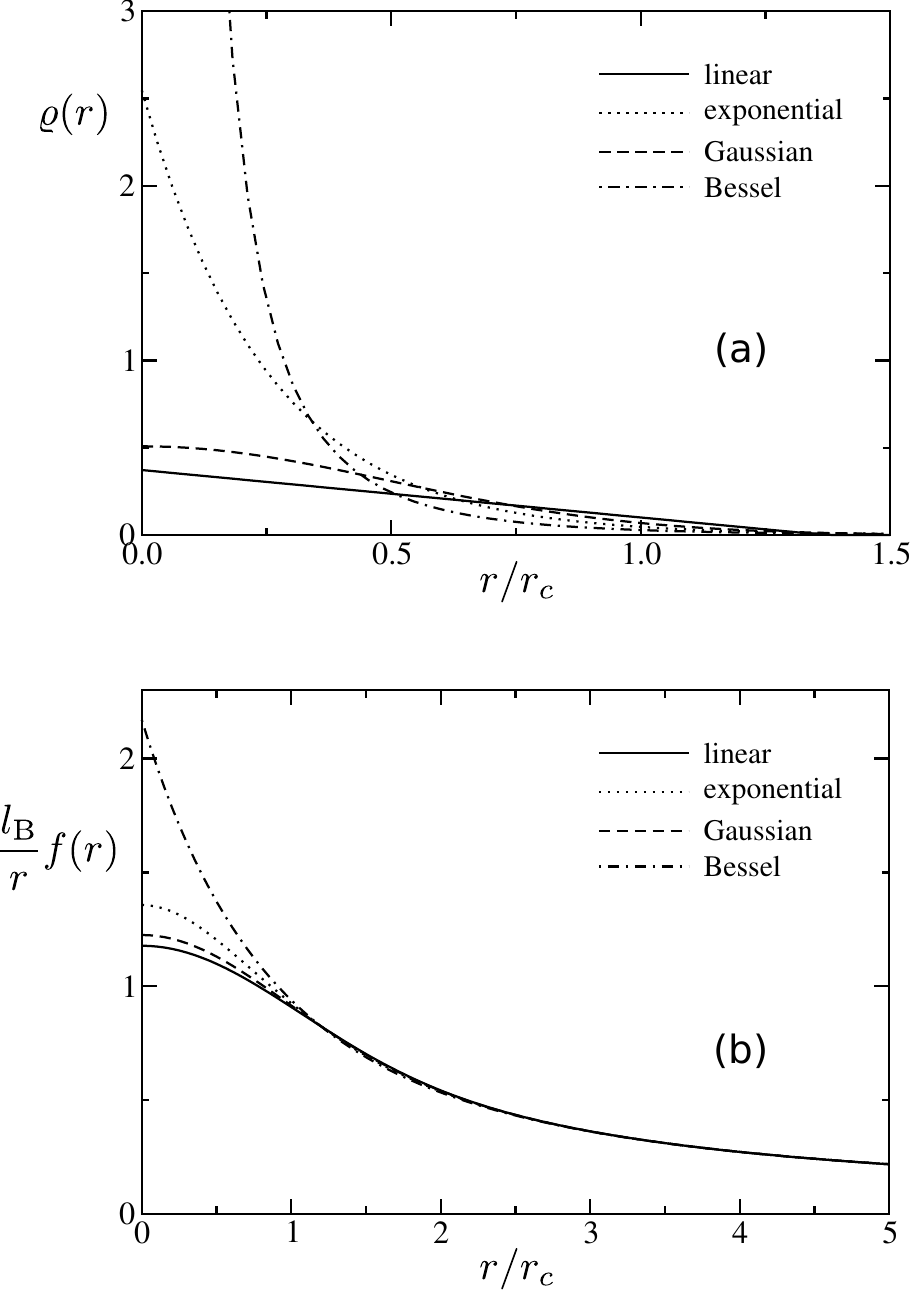}
\end{center}
\vskip -0.5cm
\caption{Charge density distributions (upper plot), and electrostatic
  interaction potentials (lower plot), for the four models in Table
  \ref{T1}.  Parameters are $\sigma=\lambda=0.5\,r_c$, $R=1.37\,r_c$
  (\ie\ matching $\rsq=0.75\,r_c^2$), and $\lB=1.1\,r_c$. The
  overlap energy $\beta u_0$ is the contact value at $r\to0$ in the
  lower plot.\label{fig:rhouplot}}
\end{figure}

The second term in \Eqref{eq:u} is the long range electrostatic
interaction.  The overall magnitude is set by the Bjerrum length
$\lB$, and the differences between the various smoothing methods
captured by a generic smoothing function $f(r)$.  We expect to recover
the Coulomb law for large separations, $U_{\mu\nu}(r)\to z_\mu z_\nu
\lB\kT/r$ for $r\to\infty$, and we also expect $U_{\mu\nu}(r)\to
\phi(0)\kT+z_\mu z_\nu u_0$ as $r\to0$, where $u_0$ is the electrostatic
overlap energy between unit charges.  This latter property ensures
thermodynamic stability, as mentioned in the introduction.  These
expectations mean that the smoothing function has the properties
$f(r)\to1$ for $r\to\infty$ and $f(r)\to\alpha r$ as $r\to0$, where
$\alpha=\beta u_0/\lB$.  Table \ref{T1} shows the charge 
density distributions and the corresponding functions $f(r)$, for the
four smoothing methods under consideration.  Fig.~\ref{fig:rhouplot}
shows representative plots for the charge densities, and corresponding
Coulomb laws, using parameter values appropriate to a 1:1 aqueous
electrolyte (see next).

Before continuing the analysis, we note that a feature common to all
the models is the existence of an additional length scale which
characterises the size (\ie\ the spatial extent) of the smoothed
charges.  This length scale (see Table~\ref{T1}) is $R$ for linear
smoothing, $\lambda$ for exponential smoothing, and $\sigma$ for
Gaussian or Bessel smoothing.  Hence, the potential defined in
\Eqref{eq:u} contains \emph{three} length scales: the DPD particle
size $r_c$, the Bjerrum length $\lB$, and separately the size of the
charge clouds.  The DPD particle size is conventionally used to
non-dimensionalise the densities, with $\rho r_c^3=3$ being widely
adopted.  Making this assumption, $r_c$ and $\lB$ are fixed by
physical arguments and the mapping to the underlying physical system,
for example $r_c=0.645\,\nm$ and $\lB=0.7\,\nm$ for a room temperature
1:1 aqueous electrolyte \cite{WVA+13}.  The ion density is then also set
by the mapping to the underlying system, for example $\rhoz
r_c^3=0.32$ for a $1\,\M$ solution \cite{WVA+13}.  These considerations
thus far fix everything except the size of the charge clouds, which is
the central issue.  In the remainder of the present study we shall
fix the values of $r_c$ and $\lB$ to correspond to this
standard mapping to a 1:1 aqueous electrolyte.

In the model it is conventional to set $\kT=1$, with the true
temperature dependence being carried by parameters in the interaction
potential, through the physical mapping.  However we note that in
terms of electrostatics, the Bjerrum length $\lB$ is a coupling
constant which when non-dimensionalised with the charge cloud size 
plays the role of an inverse temperature.  For example,
for the URPM \cite{CHK11a, CHK11b}, pronounced clustering occurs for
$\sigma/\lB\alt0.03$ and a condensation transition for
$\sigma/\lB\alt0.01$. In general we shall find similar effects in
\emph{any} of the models, whenever the size of the charge clouds is
too small.  For some applications, for example to strongly correlated
ionic systems, clustering and phase separation are actually desirable
since they can be tuned to represent real physics \cite{Lev02}.  For the
present situation though (\eg\ $\alt0.1\,\M$ electrolytes),
clustering and phase separation are physical manifestations of the
loss of thermodynamic stability as $\sigma\to0$, and as such are
unwanted low temperature artefacts.  In practice such
artefacts are not usually an issue since there is a strong incentive
to make charge clouds as large as possible, to reduce the cost of
computing the electrostatic interactions (to a specified accuracy) in
a numerical simulation.

Now we return to the analysis.  The Fourier transform of the pair
potential is given by 
\begin{equation}
\beta \tilde U_{\mu\nu}(k)=\tilde\phi(k)+z_\mu z_\nu \frac{4\pi\lB}{k^2}\,g(k)
\label{eq:ut}
\end{equation}
where $\tilde \phi(k)=\int\!d^3\rvec\, e^{-i\kvec\cdot\rvec}
\,\phi(r)$. 

Here $g(k) = k \,{\textstyle\int_0^\infty}\!dr\, \sin(kr)\,f(r)$ is
essentially a sine transform of $f(r)$.  The inverse transform is $f(r) =
({2}/{\pi})\,{\textstyle\int_0^\infty}\!dk\, \sin(kr)\,{g(k)}/{k}$.
As we shall see, the function $g(k)$ is the `glue' which ties together
all the subsequent results.  To discover its physical meaning,
consider the Coulomb interaction between a pair of identical charge
clouds $\varrho(\rvec)$, of unit magnitude
($\int\!d^3\rvec\,\varrho(r)=1$),
\begin{equation}
\beta U(\rvec)=\lB\int\!d^3\rvec_1\,d^3\rvec_2
\frac{\varrho(\rvec_1)\,\varrho(\rvec_2-\rvec)}
{|\rvec_2-\rvec_1|}\,.
\end{equation}
This takes the form of a double convolution, therefore in reciprocal
space $\beta \tilde U= {4\pi\lB\tilde\varrho^2}/{k^2}$, and we identify
$g=\tilde\varrho^2$.  This is the route used in Table \ref{T1} to
calculate the functions $g(k)$ and $f(r)$ from the charge density
$\varrho(r)$.  For completeness note that the three dimensional
Fourier transforms reduce  to 
%
$\tilde\varrho(k) = ({4\pi}/{k}) \,{\textstyle\int_0^\infty}\!dr\,
\sin(kr)\,r\,\varrho(r)$ 
%
and
%
$\varrho(r) = ({2\pi^2r})^{-1}{\textstyle \int_0^\infty}\!dk\,
\sin(kr)\,k\,\tilde\varrho(k)$, 
%
by virtue of radial symmetry.

An implication of $g=\tilde\varrho^2$ is that $g\ge0$. In fact this
provides a necessary and sufficient condition for the interaction
potential to correspond to the interaction between (identical) charge
clouds.  For example the obvious truncation $\beta U=\beta
u_0=\lB/r_c$ for $r<r_c$ and $\beta U=\lB/r$ for $r\ge r_c$ gives rise
to $g(k)$ which oscillates in sign, and therefore does not correspond
to the interaction between charge clouds (this does not necessarily
preclude the use of this potential of course).

Expanding 
the Fourier representation of $\tilde\varrho(k)$ gives
\begin{equation}
\tilde\varrho(k)=1-\smallfrac{1}{6}k^2\rsq+O(k^4)
\end{equation}
where $\rsq={\int d^3\rvec\, r^2\varrho}$ is the second moment of the
charge distribution.  This expansion implies
\begin{equation}
g(k)=1-\smallfrac{1}{3}k^2\rsq+O(k^4)\,.\label{eq:gkr2}
\end{equation}
This allows us to extract the second moment of the charge distribution
from $g(k)$.  

Now consider the overlap energy.  We have
\begin{equation}
\begin{split}
\frac{\beta u_0}{\lB}=\lim_{r\to0}\frac{f(r)}{r}
&=\lim_{r\to0}\frac{2}{\pi}
\int_0^\infty\! \!dk\,\frac{\sin (kr)}{kr}g(k)\\
&{}\hspace{6em}=({2}/{\pi})
{\textstyle \int_0^\infty}\!dk\,g(k)\,.
\end{split}\label{eq:u0calc}
\end{equation}
Again we have a result expressed in terms of the function $g(k)$.  The
final integral in \Eqref{eq:u0calc} may be tractable even if the
inverse sine transform is not.

Table~\ref{T1} shows these properties calculated for the four smoothed
charge models considered in this study.  

\section{Screening properties}
We now describe the screening properties of the four charge smoothing
methods, in the context of the above mesoscale electrolyte model.  The
details are much the same as already presented for the Gaussian
case \cite{WVA+13}.  We shall establish the conditions under which the
screening length is real-valued and the extent to which it matches
the expected Debye length in the physical system.

For the first step, the electrolyte model just described is
structurally characterised by the pair distribution functions
$g_{\mu\nu}(r)$. From these we define the total correlation
functions $h_{\mu\nu}=g_{\mu\nu}-1$.  At low densities and weak
coupling (\ie\ $\lB$ small compared to the charge cloud size) the
total correlation functions for the ionic species show a universal
exponential decay at large distances, $h_{\pm\pm} \sim e^{-\kappa
  r}/r$ for $r\to\infty$, where $\kappa^{-1}$ is the (real-valued)
screening length.  On the other hand, at high densities and/or strong
coupling, the functions $h_{\pm\pm}$ show damped oscillatory decay and
the screening length becomes complex \cite{JN09}.  The two
behaviours are separated by a sharp transition as a function of
density and coupling strength, known as the Kirkwood line \cite{Kir36}.

\begin{figure}
\begin{center}
\includegraphics[clip=true,width=2.6in]{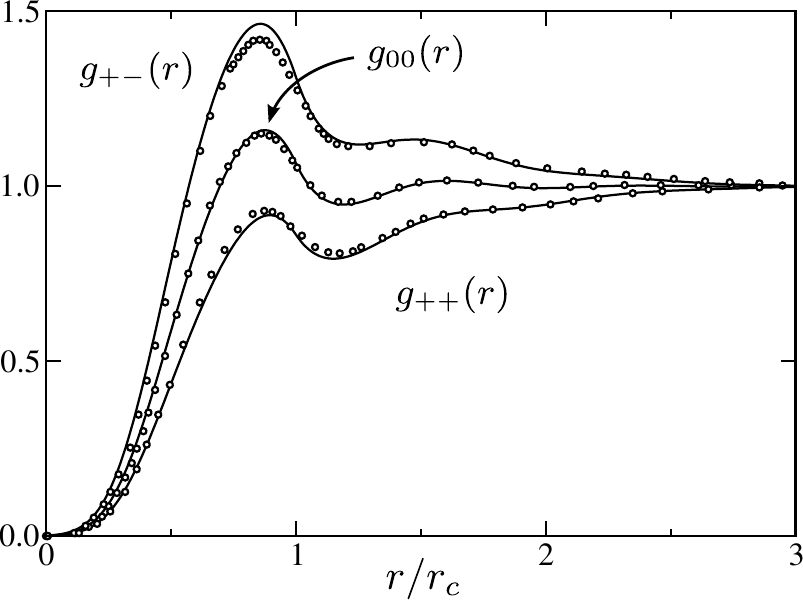}
\end{center}
\vskip -0.5cm
\caption{Comparison between HNC (solid lines) and simulation (circles,
  taken from Fig.~3 of Ref.~\onlinecite{Gro03}) for an electrolyte
  model with linear charge smoothing. Parameters are $|z_\pm|=1$,
  $\lB=1.1\,r_c$, $R=1.6\,r_c$, $\rho r_c^3=3$, and $\rho_zr_c^3=0.2$;
  corresponding to a $0.6\,\M$ 1:1 aqueous electrolyte.  The short range
  potential is $\phi(r)=\half A(1-r/r_c)^2$, with
  $A=25$.\label{fig:rdgcomp}}
\end{figure}

\begin{figure}
\begin{center}
\includegraphics[clip=true,width=3.0in]{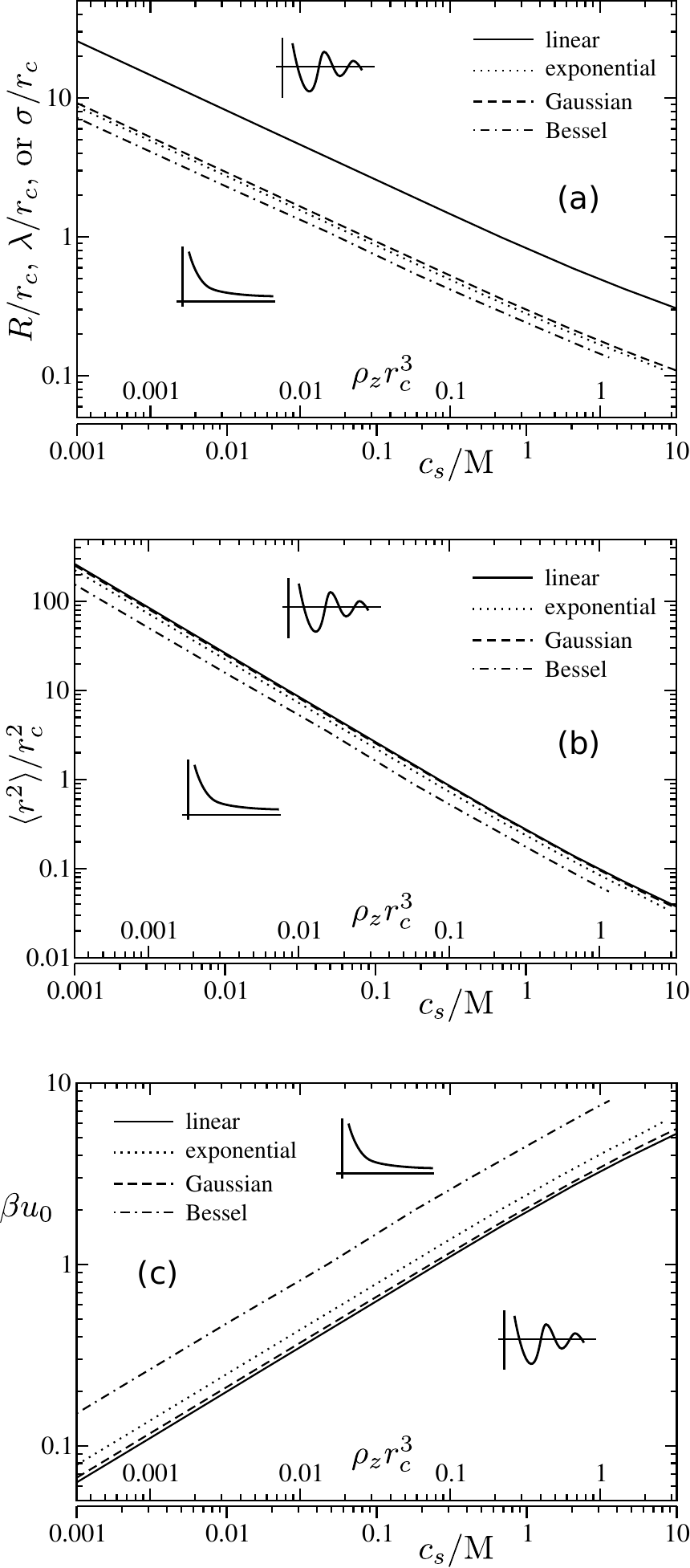}
\end{center}
\vskip -0.5cm
\caption{Kirkwood lines in various representations (see text for
  details).  The schematic insets indicate on which side of the
  Kirkwood line the total correlation functions $h_{\pm\pm}(r)$
  exhibit pure exponential or damped oscillatory asymptotic
  decay.\label{fig:genkl}}
\end{figure}

\subsection{Hypernetted chain (HNC) closure}\label{sec:hnc}
We access the total correlation functions by solving the
multicomponent hypernetted chain (HNC) integral equation closure for
the fluid.  This approximation has been extensively discussed in the
existing literature \cite{HM06, VLK+09}, and is known to be quite
accurate for these charged soft sphere models \cite{CHK11b, WVA+13}.
For example, for linear screening, Fig.~\ref{fig:rdgcomp} shows the
HNC pair distribution functions for the indicated state point compared
to simulation results from Ref.~\onlinecite{Gro03}.  The agreement is
quite good.  Other tests are reported in Ref.~\onlinecite{WVA+13}, for
the Gaussian charge model.  Our HNC code uses potential
splitting methods \cite{SPSJ73, Ng74}, and the only modification needed
to the code used in Ref.~\onlinecite{WVA+13} is to swap in the
generalised $g(k)$ functions in Table~\ref{T1}.  In particular, it is
not necessary to know the electrostatic pair potential in real space
and $f(r)$ is not required.

The HNC Kirkwood lines for the four smoothed charge models are shown
in Fig.~\ref{fig:genkl}, in three different representations.  They are
found by visually inspecting the asymptotic behaviour of the HNC total
correlation functions.  We use the iterative procedure outlined in
Ref.~\onlinecite{WVA+13}, stopping when the Kirkwood line has been
located to around 1\% accuracy in density.  We have found that the
Kirkwood lines and the screening behaviour in general is very
insensitive (\ie\ changes by at most by 1--2\%) to the presence of the
neutral solvent species and the short range repulsion.  This is very
helpful as it sharply reduces the complexity of the problem.

We note that the numerical solution method for HNC fails if the charge
size becomes too small, typically less than 10\% of $\lB$.  This loss
of solution has also been observed by Coslovich, Hansen and Kahl for
the URPM \cite{CHK11b}, and is almost certainly indicative of a
mathematical property of HNC rather than a numerical problem.

The Kirkwood lines are plotted as functions of the ion density, using
the native parameters (Fig.~\ref{fig:genkl}a), the second moment of
the charge distribution (Fig.~\ref{fig:genkl}b), and the overlap
energy (Fig.~\ref{fig:genkl}c).  We see that the second representation
brings the Kirkwood lines very close together.  This implies
Fig.~\ref{fig:genkl}b is a \emph{quasi-universal map}
which can be used as a guide for arbitrary smoothed charge models,
provided that the second moment of the charge distribution is used as
a length scale.  Further confirmation of this role for $\rsq$ and a
simple expression for the quasi-universal Kirkwood line is given in
Section \ref{sec:rpa}.  Examining the Kirkwood lines, it is obvious that
the Bessel model is something of an outlier.  In fact this is already
apparent in Fig.~\ref{fig:rhouplot} and Table~\ref{T1}, since unlike
any of the other charge distributions, $\varrho(r)$ for the Bessel
case diverges (as $1/r^2$) in the limit $r\to0$.

The horizontal axis in the three plots in Fig.~\ref{fig:genkl} is the
ion density, drawn either as a dimensionless simulation variable
$\rho_zr_c^3$ (upward pointing tick marks), or using physical units
where $c_s$ is the ion concentration in Molar units (downward pointing
tick marks).  The two are related by a simple proportionality: given
the choice $r_c=0.645\,\nm$ one has $\rho_zr_c^3 \approx
0.32\,c_s/\,\M$.

At this point we should inject a note of common sense concerning
the relevance of large values of $c_s$.  In the physical system, the
actual screening length is often taken to be the Debye length, given
by the well known expression $\kD^{-1} = 0.31\,\nm / \sqrt{c_s/\M}$,
familiar from the field of colloid science \cite{VO48}.  The Debye
length is a decreasing function of $c_s$, and $\kD^{-1}\alt r_c$
occurs for $c_s\agt0.23\,\M$.  When the Debye length falls below the
DPD particle size, it no longer makes sense to represent the behaviour
by a sophisticated electrostatic model, since the charge interactions
could equally be captured in the short range DPD potential.  In any
case, deviations from Debye-H\"uckel theory start to become
significant for $c_s\agt0.1\,\M$ and specific ion effects become more
and more important.  These considerations mean that
$c_s\approx0.1$--$0.2\,\M$ is a natural upper limit for the attempt to
match the screening properties of the mesoscale model to the physical
system.

In terms of the maps in Fig.~\ref{fig:genkl}, this implies the
screening behaviour is moot for such high values of $c_s$.  Turning
this around, we can use this to propose sensible limits for the
parameters in the models.  For example, reading from
Fig.~\ref{fig:genkl}a, the requirement to remain on the `right' side
of the Kirkwood line (\ie\ on the low density side where the model has
a real-valued screening length) translates into natural upper bounds
for $R$, $\lambda$, and $\sigma$, in units of $r_c$.
Fig.~\ref{fig:genkl}b is most useful in this respect since (setting
aside the Bessel model which we already acknowledge is an outlier)
there is a quasi-universal Kirkwood line.  Sensible screening
behaviour for $c_s\alt0.1$--$0.2\,\M$ corresponds to
$\rsq\alt(1.3$--$2.5)\,r_c^2$.  These bounds can be tightened
still further by considering the behavior of the actual screening
length---see Section \ref{sec:disc}.

Fig.~\ref{fig:genkl}c is less useful in this respect since the
relevant Kirkwood lines are not as closely collapsed together as
Fig.~\ref{fig:genkl}b.  Nevertheless the map indicates one should
ensure $\beta u_0\agt0.6$--0.9.  It is worth remarking that
heuristic considerations led Groot to propose $\beta u_0\approx1$ as
the criterion for choosing $R$ in the linear smoothing
model \cite{Gro03}, and this was later taken over to the exponential
smoothing case by Gonz\'alez-Melchor \etal\ \cite{GMV+06}.  This is in
accord with the present analysis.  Note however one should not
increase $\beta u_0$ too much (equivalent to shrinking the charge
cloud size), since that would lead towards the aforementioned low
temperature artefacts (clustering, phase separation, \etc).

\subsection{Random phase approximation (RPA)}
\label{sec:rpa}
Another well-trodden approach, also known to be quite accurate for
these charged soft sphere models, is the random phase approximation
(RPA).  In this case an analytic solution for the Fourier-transformed
total correlation functions can be obtained.  The solution method is
described in Ref.~\onlinecite{WVA+13}.  The general result is
\begin{equation}
\tilde h_{\mu\nu}=
-\frac{4\pi\lB z_\mu z_\nu g(k)}
{k^2+\kD^2 g(k)}
-\frac{\tilde\phi}{1+\rho\tilde\phi}\,.
\label{eq:rpagen}
\end{equation}
Again we see the relevance of the `glue' function, $g(k)$, which
together with $\tilde\phi(k)$ contains all the model-dependent
features.

In \Eqref{eq:rpagen}
\begin{equation}
\kD^2=4\pi\lB\,{\textstyle\sum_\mu} z_\mu^2\,\rho_\mu\label{eq:kd}
\end{equation}
is the square of the Debye wavevector, so that $\kD^{-1}$ is the
already-introduced Debye length.  For a 1:1 electrolyte,
$\kD^2=4\pi\lB\rho_z$ (in the physical system, this gives rise to the
expression used earlier from the colloid literature).

The asymptotic behaviour of $h_{\pm\pm}(r)$ is determined by the
positions of the poles of $\tilde h_{\pm\pm}(k)$, regarded as analytic
functions in the complex $k$-plane \cite{ELdC+94, HAE06, NHK12}.  The
functions share a common set of poles.  There are two typical
scenarios.  If the nearest pole to the real axis is purely imaginary,
the asymptotic behaviour of $h_{\pm\pm}(r)$ is purely exponential and
there is a real-valued screening length set by the distance of the
pole from the real axis.  Alternatively, if the nearest poles to the
real axis are complex, the asymptotic behaviour of $h_{\pm\pm}(r)$ is
damped oscillatory and the screening length is complex.  The Kirkwood
line is determined by the crossover between the two scenarios.

In the case of the RPA, there are two sets of poles arising from the
two contributions to $\tilde h_{\pm\pm}$ in \Eqref{eq:rpagen}.  Of
these, the poles from the second term (arising from the short range
repulsion) are usually too distant from the real axis to be relevant;
therefore we focus attention on the first term (arising from the
electrostatics).  The poles in this term correspond to the zeros of
\begin{equation}
k^2+\kD^2 \, g(k)=0\,.\label{eq:rpakl}
\end{equation}
Since $\kD\to0$ at low densities, this equation can be solved
iteratively, using \Eqref{eq:gkr2} for the expansion of $g(k)$ about
$k=0$.  We find that the relevant zero is purely imaginary and the
corresponding screening length is given by
\begin{equation}
\kappa^{-1}=\kD^{-1}[1-\smallfrac{1}{6}\rsq\kD^2+O(\kD^4)]\,.
\label{eq:kser}
\end{equation}
Thus we see that, as the ion density decreases, the screening length
approaches the Debye length, from below, by an amount controlled
by $\rsq$.  This is lends weight to the argument that $\rsq$
is a good choice for matching between smoothing methods, when it comes
to predicting the screening properties.

For the Gaussian and Bessel cases, the zeros of \Eqref{eq:rpakl} can
be obtained analytically \cite{NHK12, WM13, WVA+13}.  The corresponding
Kirkwood lines lie at $\kD\sigma=e^{-1/2}$ for the Gaussian case, and
$\kD\sigma=\half$ for the Bessel case.  In fact for
$\sigma/r_c\agt0.3$ these are practically indistinguishable from the
HNC result (see also Fig.~5 in Ref.~\onlinecite{WVA+13}).  This means
that the quasi-universal Kirkwood line in Fig.~\ref{fig:genkl}b is
given by (Gaussian case)
\begin{equation}
\frac{\rsq}{r_c^2}=\frac{3}{4\pi e\lB\rho_zr_c^2}\approx
\frac{0.081}{\rho_z r_c^3}\approx\frac{0.253\,\M}{c_s}\,.
\end{equation}
Generally, for the RPA, we can infer from
\Eqref{eq:rpakl} that the Kirkwood line corresponds to some
particular value of $\kD$, measured in units of the size of
the charge clouds.  This is because the latter length scale is the
only one available to non-dimensionalise the argument of $g(k)$ (for
examples, see Table \ref{T1}).  Since $\kD\propto1/\sqrt{\rho_z}$, and
we have seen that the RPA is quite accurate, this observation explains
the slopes of the Kirkwood lines in Fig.~\ref{fig:genkl}.  

Another point to note from \Eqref{eq:rpagen} is that if the
screening properties are governed solely by the first term, the second
term can be neglected.  This implies that in the RPA the screening
properties are \emph{completely} insensitive to the presence of the
neutral solvent species and short range repulsions.  This explains the
similar observation made for the HNC.

\section{Final recommendations}
\label{sec:disc}
The charge cloud size arising from the smoothing operation is not
supposed to have a physical significance and can be chosen to
maximise computational efficiency.  For example, a large amount of
smoothing means that one can use a coarse mesh spacing in a P$^3$M
method, or fewer terms in an Ewald sum in other methods.  But as we
have seen this is in direct competition with the desire to avoid
unwanted artefacts, such as damped oscillatory behaviour in the total
correlation functions which arises when one is on the `wrong' side of
the Kirkwood line.  Clearly then, the incentive is to push the charge
cloud size towards the upper bounds indicated in the discussion in
Section \ref{sec:hnc}.  

Specific requirements for the screening behaviour can then be used to
sharpen these bounds.  Thus in Ref.~\onlinecite{WVA+13} we suggested
$\sigma= 0.5\,r_c$ for the Gaussian case, since this means that
there is a real-valued screening length, within 10\% of the Debye
length, for 1:1 electrolyte solutions with $c_s\alt0.15\,\M$.  We can
translate this into suggested parameter values for other smoothing
models by matching the second moment of the charge distribution,
namely $\rsq=0.75\,r_c^2$.  Using the final column in Table
\ref{T1}, this corresponds to the choices
\begin{equation}
\sigma=\lambda=0.5\,r_c\,,\quad\text{or}\quad R\approx1.37\,r_c\,.
\label{eq:parset}
\end{equation}
When we compute the screening length for a $c_s=0.15\,\M$ 1:1
electrolyte with these parameters, we find that all the methods give a
value within 1\% of $\kappa^{-1}\approx1.11\,r_c\approx 0.72\,\nm$.
This can be compared to
$\kD^{-1}\approx0.31\,\nm/\sqrt{c_s}\approx0.80\,\nm$.  Thus the
parameters in \Eqref{eq:parset} are not only \emph{self consistent},
but lead to a screening length \emph{within 10\%} of the Debye
length \cite{Cnote}.  Furthermore, the agreement will improve as $c_s$
decreases.  \Eqref{eq:parset} thus represents our recommended choice
of parameters for the available mesoscale electrolyte models used in
DPD.  For comparison, Groot recommended $R=1.6\,r_c$ for the linear
smoothing model, and Gonz\'alez-Melchor \etal\ recommended
$\lambda=1.08\,r_c$ for the exponential smoothing model.  Thus our own
recommendations are somewhat more stringent but come with guaranteed
behaviour in terms of the screening properties.

Although our analysis has been confined to 1:1 aqueous electrolytes,
it is not necessarily that restrictive since the methods used here can
in principle be transferred to other situations, such as concentrated
or multivalent electrolytes, and non-aqueous solvents.  An ultimate
goal is to incorporate specific ion effects into the models, such as
the Hofmeister series \cite{CW85}.  Mindful of this, and the utility of
a fast, accurate, multicomponent integral equation code in general, we
have made available the FORTRAN 90 source code used in these
calculations as fully documented open source software \cite{OZnote}.

We thank Lucian Anton and Andrew Masters for their generous advice and
guidance.  In particular we acknowledge that the computer code used
here was based on a HNC code originally developed by Lucian.  We
additionally thank Andrew for hosting visits of one of us (AV) and
providing a stimulating environment in which the work was done.  We
would also like to acknowledge the contribution of Ming Li, who noted
the structure of the RPA solution for ions and neutral spheres in a
related context.  AV acknowledges partial support for trips to
the UK from grant RFBR \#13-03-01111.


\begin{thebibliography}{23}
\bibitem [{Cno()}]{Cnote}%
  \BibitemOpen
  \href@noop {} {}\bibinfo {note} {Since $1-\smallfrac{1}{6}\rsq\kD^2\approx
  0.92$, most of this difference is explained by the leading correction in
  \Eqref{eq:kser}.}\BibitemShut {Stop}%
\end{thebibliography}

%

\end{document}